\documentclass[]{aa}
\usepackage{psfig}
\usepackage{graphicx}
\def\fdeg{\hbox{$\,.\!\!^{\circ}$}}

\begin{document}
\title{The phase of the radio and X-ray pulses of PSR~B1937$+$21}
\author{G. Cusumano\inst{1}, W. Hermsen\inst{2}, M. Kramer\inst{3},
L. Kuiper\inst{2}, O. L\"ohmer\inst{4}, E. Massaro\inst{5}, T.
Mineo\inst{1}, L. Nicastro\inst{1},   \and B.W. Stappers\inst{6}}

\institute{Istituto di Astrofisica Spaziale e Fisica Cosmica -
Sezione di Palermo, CNR, Via Ugo La Malfa 153, I-90146, Palermo,
Italy \and SRON National Institute for Space Research,
Sorbonnelaan 2, 3584 CA Utrecht, The Netherlands \and University
of Manchester, Jodrell Bank Observatory, Macclesfield, Cheshire
SK11 9DL, UK  \and Max-Planck-Institut f\"ur Radioastronomie, Auf
dem H\"ugel 69, D-53121 Bonn, Germany  \and Dipartimento di
Fisica, Universit\'a La Sapienza, Piazzale A. Moro 2, I-00185,
Roma, Italy; Istituto di Astrofisica Spaziale e Fisica Cosmica -
Sezione di Roma, CNR, Via del Fosso del Cavaliere, I-00100 Roma,
Italy \and  ASTRON, Postbus 2, 7990 AA Dwingeloo, The Netherlands;
Astronomical Institute "Anton Pannekoek", University of Amsterdam,
Kruislaan 403, 1098 SJ Amsterdam, The Netherlands}

\offprints{G. Cusumano: \\giancarlo.cusumano@pa.iasf.cnr.it}
\date{Received:.; accepted:.}

\titlerunning{Timing  and spectral analysis of PSR~B1937$+$21}
\authorrunning{G. Cusumano et al.}

\abstract{ We present timing and spectral results of PSR B1937+21,
the fastest known millisecond pulsar  ($P\simeq1.56$ ms), observed
with RXTE. The pulse profile, detected up to $\sim$20 keV, shows a
double peak with the main component much stronger than the other.
The peak phase separation is $0.526\pm0.002$ and the pulsed
spectrum over the energy range 2--25 keV is well described by a
power law with a photon index equal to $1.14\pm0.07$. We find that
the X-ray pulses are closely aligned in phase with the giant
pulses observed in the radio band. This results suggest that giant
radio pulses and X-ray pulses originate in the same region of the
magnetosphere due to a high and fluctuating electron density that
occasionally emits coherently in the radio band. The X-ray events,
however, do not show any  clustering in time indicating that no
X-ray flares are produced.
 }

\maketitle

\section{Introduction}
Millisecond radio pulsars (MSPs) differ from ordinary pulsars for
their short spin period in the range 1.5--30 ms and for their
low spin-down rate of $\dot{P} <
 10^{-19}$ s s$^{-1}$. They
are presumably very old objects, with spin down ages of $\tau
\sim 10^9$--$10^{10}$ yr and low surface magnetic fields
$B\simeq 10^8$--$10^{10}$ G. MSPs are believed to be recycled by
having accreted mass and angular momentum from an evolving
companion star (Alpar et al. 1982).
About 50\% of all X-ray detected rotation-powered pulsars are MSPs
and their X-ray emission seems to have characteristics similar to
ordinary pulsars with both thermal and non-thermal components. The
former can be described by a modified black body generally peaking
in the soft X-ray range, while the latter is characterized by a
power-law like spectrum over a broad energy band.
Data available up to now, however, leave open the debate about the
actual presence of thermal emission from MSPs. The spectra of the
few MSPs for which thermal models have been claimed, can also be
modeled either with curved or broken power laws (Becker \&
Aschenbach 2002).

PSR B1937$+$21 was the first MSP discovered (Backer et al. 1982)
and, with the period of 1.56 ms, it remains the most rapidly
rotating neutron star presently known. The distance estimated from
the observed dispersion measure (DM) and from a model for the
Galactic free electron distribution  (Taylor \& Cordes 1993,
Cordes \& Lazio 2002) is 3.6 kpc (see also discussion in Nicastro
et al. 2003). Its spin down luminosity is $\dot{E} \sim 1.1 \times
10^{36}$ erg s$^{-1}$ and the dipolar magnetic field component at
the star surface is  $\sim 4.1\times 10^{8}$ G. Like the Crab
pulsar (Lundgren et al. 1995), PSR B0540$-$69 (Johnston \& Romani
2003) and the other MSP PSR B1821$-$24 (Romani \& Johnston),
 PSR B1937$+$21 exhibits
sporadic emission of giant pulses in the radio band (Sallmen \&
Backer 1995; Cognard et al. 1996, Kinkhabwala et al. 2000). Such
pulses are extremely short events ($\tau < 0.3$ $\mu$s at 2.38
GHz) confined to small phase windows trailing the main and
interpulse.

\indent X-ray emission from this pulsar was detected by ASCA
(Takahashi et al. 2001) above 2 keV, with a pulse profile
characterized by a single sharp peak and a pulsed fraction of
44\%. Comparing the X-ray and radio phase arrival times, these
authors claimed that the X-ray pulse is aligned with the radio
interpulse.  Later, BeppoSAX detected pulsed emission from PSR
B1937+21 (Nicastro et al. 2002, 2003) and the pulse profile was
found to show a double peak pattern with a phase separation from
P1 to P2 of $0.52\pm0.02$ and a significance of the second peak of
$\sim 5\sigma$. The BeppoSAX data did not allow to study the
relative alignment between X-ray and radio pulses, because the
timing did not maintain the necessary accuracy to UTC. \\ \indent
In this letter we present the results of the timing and spectral
analysis of  a RXTE observation of PSR B1937+21. We compare the
absolute phases of the X-ray and radio pulsed signals  and show
that the X-ray peaks are phase aligned  with the radio giant
pulses (Section 3). The pulsed spectrum is derived in Section 4.

\section{X-ray and Radio Observations and Data Reduction}

The RXTE pointings at PSR B1937+21 were performed between February
22 and February 27, 2002. The total exposure times were about
140,000 s for the PCA units 0, 2 and 3, and about 20,000 s for the
units 1 and 4. Standard selection criteria were applied to the
observation data excluding time intervals corresponding to South
Atlantic Anomaly passages and when the Earth's limb was closer
than 10 degrees and the angular distance between the source
position and the pointing direction of the satellite was larger
than 0\fdeg02.  We verified that the selection of all PCA detector
layers, instead of those from the top layer only, increased
significantly the S/N ratio of the pulsation and adopted this
choice for our timing analysis. We used only data obtained with
the PCA (Jahoda et al. 1996) accumulated in ``Good Xenon''
telemetry mode for the timing and spectral analysis. Events are
time-tagged with a 1 $\mu$s accuracy with respect to the
spacecraft clock and with an absolute time accuracy of 5--8 $\mu$s
with respect to UTC. The UTC arrival times of all selected X-ray
events were first converted to the Solar System Barycentre using
the (J2000) pulsar position given in Table 1 and the JPL2000
planetary ephemeris (DE200, Standish 1982).

The radio ephemeris of PSR B1937+21 were obtained from high
precision timing observations made with the 100-m Effelsberg
radiotelescope in Bonn, Germany, and with the Westerbork Synthesis
Radio Telescope (WSRT) in Westerbork, The Netherlands. From
October 1996 timing data were collected at 1410 MHz with the
Effelsberg telescope once per month, with a typical observing time
of 3$\times$7 min. Two circular polarized signals were mixed down
to an intermediate frequency, detected and coherently de-dispersed
in a digital filterbank (Backer et al. 1997).  The data were time
stamped with clock information from a local hydrogen maser clock
and later synchronized to UTC(NIST) using the signals from the
Global Positioning System (GPS). Pulse times-of-arrival (TOAs)
were calculated from a fit of a synthetic template to the observed
profile with a template matching procedure (for details see Lange
et al.\ 2001). TOA errors for PSR~B1937+21 lie in the range
80--200 ns making it one of the most precise objects for pulsar
timing.

The WRST observations were performed since July 1999
 at center frequencies of 840 and 1380 MHz using the Dutch pulsar
 machine PuMa (Vo\^ute et al. 2002). The observing and analysis procedure
 to produce the WSRT TOAs were similar to those at Effelsberg except that linear
 polarisations were recorded. The dual frequency nature of the WSRT data set
 allowed us to accurately monitor dispersion measure (DM) variations which could cause
 significant uncertainties in the absolute timing of pulsars
 (Backer et al.\ 1993\nocite{bhh93}).

In the timing analysis, both sets of TOAs obtained at Effelsberg
and
 WSRT were first independently fit to a pulsar spin-down model with
 the software package {\sc
 tempo}\footnote{http://pulsar.princeton.edu/tempo}. The resultant radio
 ephemerides were then used for aligning the RXTE data with the radio
 profiles, producing fully compatible results. Finally, we produced a
 best-fit timing model for PSR B1937+21, as given in Table 1, from the
 combined Effelsberg and WSRT TOAs to align the RXTE and radio data
 (see Fig. 1).

{\bf
\begin{table} \caption{JPL DE200 ephemeris of PSR B1937+21
(data from Effelsberg and WSRT observations).
 \label{tabeph}}
\begin{flushleft}
\begin{tabular}{ll}
\hline \noalign{\smallskip} {\bf Parameter} &  {\bf Value}
 \\
\hline \noalign{\smallskip}
 Right Ascension (J2000) & $19^{\rm h}$
$39^{\rm m}$ 38\fs560096(6)    \\
 Declination (J2000) &
$21^\circ$ $34'$ 59\farcs13552(13) \\
 Proper motion in R.A. (mas yr$^{-1}$) & $-0.05(2)$ \\
 Proper motion in Dec (mas yr$^{-1}$) & $-0.47(4)$ \\
 Frequency (Hz) & 641.928244534462(3) \\
 Frequency derivative (Hz s$^{-1}$) & $-4.331046(13)\times 10^{-14}$ \\
 Frequency 2nd derivative (Hz s$^{-2}$) & $8.9(1.8)\times 10^{-27}$ \\
 Epoch of the period (MJD) & 52328.0 \\ DM (cm$^{-3}$ pc) & 71.02666(16) \\
 DM derivative (pc cm$^{-3}$ yr$^{-1}$) & $-2.10(17)$  \\
 Epoch validity start/end (MJD) & 50360 -- 52867 \\
 Epoch of arrival time$^\dagger$ (MJD) & 52332.167314839268(11) \\
 Frequency of arrival time (MHz) & 1409.300  \\
 Post-fit timing rms (ns) & 910 \\
 \noalign{\smallskip} \hline
\end{tabular}
Note: Uncertainties quoted are in the last digit(s) and represent
$2\sigma$ estimates (twice the formal {\sc tempo} errors).  The
Epoch of arrival time refers to the maximum of the main peak in
the integrated radio profile.\\
 $^\dagger$ at Effelsberg telescope
\end{flushleft}
\end{table}
}

\section{Pulse profile and phase analysis}

RXTE data were searched for pulsed emission
by using the folding technique in a range centered at a value
computed from the ephemeris reported in Table 1. The plot of the
$\chi^2$ vs the pulsar frequency showed a clear single maximum,
very prominent above the noise level and the pulsar frequency,
estimated by fitting the $\chi^2$ peak with a gaussian profile,
was $\nu=641.92824453\pm 2\times 10^{-8}$ Hz in agreement
within the errors with that  from the radio timing model (Table
1).

\begin{figure}
\centerline{ \vbox{ \psfig{figure=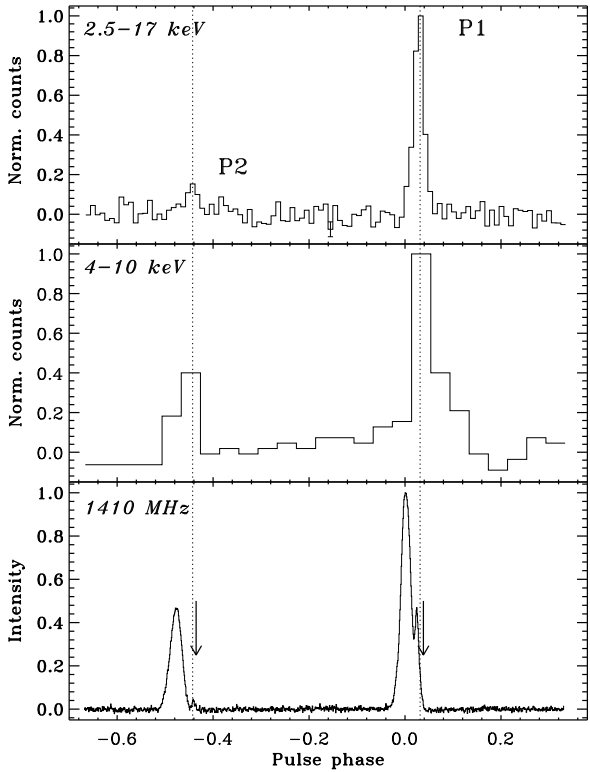,width=8.0cm,angle=0}
}} \caption{ {\bf Top:} The RXTE pulse profile of PSR B1937+21 in
the 2--17 keV energy band. The bin size corresponds to $\sim$ 16
$\mu$s. {\bf Middle:} The 4--10 keV BeppoSAX profile (Nicastro et
al. 2002) with the P1 aligned with the P1 phase in the top panel.
{\bf Bottom:} Radio pulse profile at 1.6  GHz obtained from
Effelsberg. Vertical arrows indicate the phases of observed radio
giant pulses. }
\end{figure}

The highest significance of the pulsation is reached in the energy
interval 2.5--17.0 keV and the resulting X-ray pulse profile,
obtained by folding the data with the radio frequency (Table 1),
is shown in Fig. 1 (top panel). It is characterized by a prominent
narrow first peak (P1), and by a less apparent second peak (P2).
The significance of the latter is $5\sigma$ above the off-pulse
level. P2 is lagging P1 by $0.526\pm0.002$, determined by fitting
both pulses with symmetric Lorentzian shapes. The detection of P2
is confirmed by the BeppoSAX data in which a second peak was
detected at the same phase location (Nicastro et al. 2002). We
show the BeppoSAX result in the middle panel of Fig. 1: this
profile has been shifted to align P1 with the phase of P1 in the
top panel. Pulse widths are wider in the BeppoSAX data likely
because the events are affected by a less accurate time tagging.
The P1 width (FWHM) measured in the RXTE profile is only $29 \pm
2$ $\mu$s and the P2 width is $51 \pm 21$ $\mu$s. Fig. 1 (bottom
panel) shows the radio profile from one Effelsberg observation.
Vertical arrows mark the phases of giant pulses (Kinkhabwala et
al. 2000). The comparison in absolute phase between the X-ray and
radio profiles shows that the P1 lags the main radio pulse by $44
\pm 1 \pm 5 \pm 8$ $\mu$s and P2 lags the secondary radio peak by
$51 \pm 3 \pm 5 \pm 8$ $\mu$s, where the quoted uncertainties are
due to statistical error, source position inaccuracies and
absolute timing precision of RXTE, respectively. The X-ray peaks
appear closely aligned with the phase of the radio giant pulses.
In addition, the phase separation between the X-ray pulses of
$0.526 \pm 0.002$ is more consistent with the phase separation
between the positions of the giant radio pulses ($0.5264 \pm
0.0006$) than with that between the radio main and secondary
pulses ($0.52106 \pm 0.00003$). The latter makes a systematic
difference in the absolute X-ray and radio timing as explanation
for the shifts unlikely.
\\
\indent The occurrence of the same phases for the X-ray pulses and
the radio giant ones suggests the possibility that high energy
photons are emitted simultaneous with the latters.
Therefore, we searched  if there is some evidence for a bunching
of X-ray photons with a rate similar to that of giant pulses and
equal to $\sim4$ pulses per minute (Cognard et al. 1996;
Kinkhabwala et al. 2000). During the RXTE exposure we then
expected that pulsed events occur in about 9000 X-ray flares. To
investigate this hypothesis we made an X-ray light curve selecting
only events within the phase interval centered in P1 with a phase
width of $\Delta \phi = 0.06$ (90 $\mu$s) and studied the
frequency distribution of these events. Since the dead time of the
PCA is about 10 $\mu$s the maximum content of a bin in the
presence of a X-ray flares cannot exceed 8--9 counts. We found the
following statistics: 2 bins with 4 counts, 11 bins with 3 counts,
574 bins with 2 counts, 294\,060 bins with 1 count and
92\,208\,884 with 0 count. This distribution is not consistent
with the Poisson statistics, where the expected number of bins
with a number equal or higher than 2 counts is much lower than
measured. However, there is no evidence for the existence of X-ray
giant pulses because the number of bins with a content different
from the Poisson distribution was only 116, much lower than the
number foreseen from the frequency of radio giant pulses. Another
possibility is that the rate of X-ray giant pulses could be lower
than that observed in the radio band and that the high energy
emission could be a mix of steady pulsation plus some more rare
giant pulse episodes. We constructed other light curves selecting
events in 10 different phase intervals far from P1 and P2 and with
the same phase width  used in the selection of the P1 interval. We
found similar deviations from the expected Poisson distribution in
all light curves. In particular, the number of bins deviating from
a Poisson distribution was found to be between 60 and 150.
Therefore, we conclude that there is no evidence that the X-ray
emission of PSR B1937+21 is bunched in relatively rare events of
high intensity.

\section{Spectral analysis}
Pulse-phase histograms were evaluated separately for each unit of
the PCA for the 256 PHA channels. Response matrices and sensitive
areas were then derived for each PCU and summed weighting them by
the integrated background subtracted counts of the corresponding
PCU's pulse phase histograms. Pulsed spectra of the main pulse
were obtained accumulating in the phase interval 0.68--0.72 and
subtracting the mean off-pulse level evaluated in the phase
interval 0.27--0.65.
Because of the low intensity of P2, its phase interval was not
added to the pulsed signal. Spectra were finally combined by
summing the individual PHA counts and assigned a total exposure
time equal to the sum of the individual exposures. Furthermore,
energy channels were rebinned to have a minimum content of 20
counts. The resulting pulsed spectrum in the range 2--25 keV was
fitted with an absorbed power-law model with the interstellar
hydrogen column density $N_{\rm H}$ fixed to $2.0\times 10^{22}$
cm$^{-2}$ (Nicastro et al. 2003). The best fit pulsed spectrum
gave a photon index of 1.14$\pm$0.07 with a reduced $\chi^2$ of
0.90 (47 d.o.f): it is shown in Fig. 2 (top panel) with the
residuals in units of data to model ratio (bottom panel). These
results are in agreement with those obtained from the analysis of
the data of a BeppoSAX observation (Nicastro et al. 2003). The
unabsorbed flux in the 2--25 keV band is $6.6\times10^{-13}$
erg~cm$^{-2}$~s$^{-1}$ with the
corresponding luminosity $L_X = 8.4\times 10^{31}\Theta$ ($d/3.6$
kpc)$^2$ erg s$^{-1}$, where $\Theta$ is the pulsar beam size.

A fit with a black body model gives  $kT=2.6\pm0.1$ keV, with a
$\chi^2_r=1.5$ (47 d.o.f). Although the fit is marginally
acceptable, there are some systematic deviations in the residuals,
and the derived temperature $T=2.8\times10^7$ K is very high,
particularly for an old recycled MSP.


\begin{figure}
\centerline{ \vbox{
\psfig{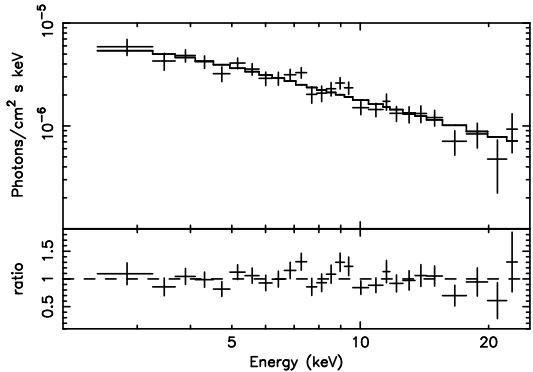} }} \caption{The
pulsed spectrum (P1) with superimposed the best fit power law
model ({\bf top}) and their ratio ({\bf bottom}) }
\end{figure}

\section{Discussion}
 The RXTE observations of the millisecond pulsar PSR B1937+21
 provided for the first time a detection of a pulsed emission up
 to an energy of 20 keV.
Like in the radio, we find a double peaked pulse shape with a
dominant first peak and a much weaker second peak. Despite being
only weakly detected, its presence is confirmed
  by recent BeppoSAX data showing a  second peak at the
  exact same location.
The pulsed spectrum is remarkably hard with a photon index of 1.14
indicating that the X-ray emission is non-thermal. The major
finding of the present analysis is that the X-ray and radio pulses
are not precisely aligned, but the formers lag the latters by a
phase difference of $0.028\pm0.006$ (P1). The X-ray pulses are
then very well aligned to the giant pulses observed in the radio
band (Kinkhabwala \& Thorsett 2000). Our result is not in
agreement with the single peak profile at the same phase of the
radio interpulse, reported by Takahashi et al. (2001) from ASCA
data.

Phase coincidences between radio giant and X-ray pulses have
earlier been reported by Romani \& Johnston (2001) for the MSP PSR
B1821$-$24. These authors predicted that PSR B1937$+$21 must also
show the same property, which has now been confirmed. Among the
ordinary pulsars, giant pulses are observed only from the Crab
pulsar (e.g. Lundgren et al. 1995, Cordes et al. 2003) and very
recently from PSR B0540$-$69 (Johnston \& Romani 2003).

The fact that X-ray and giant radio pulses are observed in the
same narrow phase intervals suggests that they are emitted from
the same region of the magnetosphere. It is indeed very unlikely
that travel time and aberration effects will combine themselves to
give exactly the same phases. It is unknown if this region lies
close to the polar caps or in some other region of the
magnetosphere. According to the scenario proposed by Romani \&
Johnston (2001) these pulses are produced in the outer gaps where
there is copious production of secondary $e^+e^-$ pairs. In this
respect it is important to note that PSR B1937+21 is the only
known pulsar with an estimated magnetic field strength at the
light cylinder larger than that of the Crab pulsar, while PSR
B1821$-$24, has a field at the light cylinder similar to the Crab.
Also PSR B0540$-$69 ranks near the top of the distribution of
pulsars with strong B fields near the light cylinder (about half
the strength of the Crab and PSR B1821$-$24). The X-ray pulsed
emission should be characterized by a stable intensity, in fact we
do not find evidence for any clustering in time that could
indicate the presence of X-ray flares.
  An interesting possibility is that there are large
instabilities affecting mainly the space distribution of secondary
pairs rather then their number. Likely, the current across the gap
should remain nearly constant. It is then possible that the
amplitude of these spatial fluctuations may occasionally be very
large to produce an enhanced coherence in the radio emission
observed as giant pulses. We expect that these fluctuations occur
over a very short distance scale, comparable or smaller than $c
\Delta \phi / \nu \simeq 1.3\times10^5$ cm, equal to a fraction of
about 4.4$\times10^{-4}$ of the radius of the velocity-of-light
cylinder, to keep the phase window of giant pulses of the order of
$\Delta \pi = 2.8\times 10^{-3}$, corresponding to 1$^{\circ}$
as found by Kinkhabwala \& Thorsett (2000).

\begin{acknowledgements}
Radio results are based on observations with the 100-m telescope
of the Max-Planck-Institut f\"ur Radioastronomie at Effelsberg.
\end{acknowledgements}

\end{document}